\documentclass[namedreferences]{solarphysics}

\usepackage[optionalrh]{spr-sola-addons} 
\usepackage{savesym}
\usepackage{graphicx}        
\usepackage{gensymb}
\usepackage{amssymb}        
\savesymbol{iint}
\savesymbol{iiint}
\usepackage{amsmath}
\usepackage{color}           
\usepackage{natbib}          
\usepackage{url}

\begin{document}

\begin{article}
\tracingmacros=2
\begin{opening}

\title{FLUKA Simulations of Pion Decay Gamma-radiation from Energetic Flare Ions}

\author[addressref={aff1},corref,email={alexander.mackinnon@glasgow.ac.uk}]{\inits{A.L.}\fnm{Alexander}~\lnm{MacKinnon}\orcid{0000-0002-3558-4806}}
\author[addressref={aff2},email={sergio.szpigel@mackenzie.br}]{\inits{S.}\fnm{Sergio}~\lnm{Szpigel}\orcid{0000-0003-2529-2225}}
\author[addressref={aff2,aff3},email={guigue@craam.mackenzie.br}]{\inits{G.}\fnm{Guillermo}~\lnm{Gimenez de Castro}\orcid{0000-0002-8979-3582}}
\author[addressref={aff2},email={tuneu.jordi@gmail.com}]{\inits{J.}\fnm{Jordi}~\lnm{Tuneu}\orcid{0000-0002-4367-7324}}


\address[id=aff1]{School of Physics and Astronomy, University of Glasgow, GLASGOW G12 8QQ, UK}
\address[id=aff2]{Centro de Radio Astronomia e Astrofisica Mackenzie, Universidade Presbiteriana Mackenzie, Sao Paulo, Brazil}
\address[id=aff3]{Instituto de Astronomia y Fisica del Espacio, UBA/CONICET, Buenos Aires, Argentina.}

\runningauthor{A. L. MacKinnon et al.}
\runningtitle{FLUKA Flare $\gamma$-ray Calculations}

\begin{abstract}
Gamma-ray continuum at $> 10 $ MeV photon energy yields information on $\gtrsim 0.2 - 0.3$ GeV/nucleon ions at the Sun. We use the general-purpose Monte Carlo code \emph{FLUktuierende KAskade} (FLUKA) to model the transport of ions injected into thick and thin target sources, the nuclear processes that give rise to pions and other secondaries and the escape of the resulting photons from the atmosphere.  We give examples of photon spectra calculated with a range of different assumptions about the primary ion velocity distribution and the source region. We show that FLUKA gives results for pion decay photon emissivity in agreement with previous treatments. Through the directionality of secondary products, as well as Compton scattering and pair production of photons prior to escaping the Sun, the predicted spectrum depends significantly on the viewing angle. Details of the photon spectrum in the $\approx 100$ MeV range may constrain the angular distribution of primary ions and the depths at which they interact. 
We display a set of thick-target spectra produced making various assumptions about the incident ion energy and angular distribution and the viewing angle. If ions are very strongly beamed downward, or ion energies do not extend much above 1 GeV/nucleon, the photon spectrum is highly insensitive to details of the ion distribution. Under the simplest assumptions, flares observed near disc centre should not display significant radiation above 1 GeV photon energy. We give an example application to Fermi Large Area Telescope data from the flare of 12 June 2010. 

\end{abstract}
\end{opening}
\tracingmacros=0

\section{Introduction}
     \label{S-Introduction} 
Gamma-rays from solar flares carry information on energetic ion populations at the Sun \citep[e.g.][]{2011SSRv..159..167V}. In particular, the ~100 MeV pion decay spectral component results from the highest energy flare ions and was first observed from solar flares with the Gamma-Ray Spectrometer (GRS) instrument on the Solar Maximum Mission (SMM) \citep{1985ICRC....4..146F}. This radiation often shows a distinctive spectral flattening around $m_{\pi^0}c^2/2 = 67$~MeV (here $m_{\pi^0}$ is the neutral pion rest mass) that is inconsistent with bremsstrahlung. It thus testifies unambigously to the presence at the Sun of accelerated ions whose energies exceed ~200 - 300 MeV/nucleon. Observations with the Fermi Large Area Telescope \citep[LAT -][]{2009ApJ...697.1071A} have shown that it occurs more frequently than previously appreciated, even in smaller M-class flares, that it can extend well into the GeV photon energy range and that it sometimes persists for many hours after all the other features of a flare have concluded \citep{2012ApJ...745..144A,2014ApJ...787...15A,2018ApJ...869..182S}.

Ions whose energies exceed the relevant thresholds can produce both charged and neutral pions. The resulting (angle-averaged) gamma-ray spectrum was discussed in detail by \cite{1987ApJS...63..721M} and its depth- and angle-dependence by \cite{1992ApJ...389..739M} and \cite{2010ApJ...721.1174T}. The neutral pions decay immediately to two photons giving a spectrum with a maximum at 67 MeV, symmetric in $\log\epsilon$ ($\epsilon$ is the photon energy), with a width determined by the energies of the primary ions \citep{1970Ap&SS...6..377S,1986A&A...157..223D}. Charged pions decay to muons and then to electrons and positrons which emit gamma-rays by bremsstrahlung as they slow down (with a few \% additional contribution from positron annihilation in flight). The relative proportions of $\pi^0$ and $\pi^\pm$ are determined once the properties of the primary ions (chemical composition, energy distribution) and of the source region (chemical composition, thick or thin) are specified but the bremsstrahlung spectral component may be reduced relative to that from $\pi^0$ decay if synchrotron losses of the e$^\pm$ are significant. Because the primary particles are positively charged, e$^-$ are significantly less numerous than e$^+$.

 The highest energy particles accelerated in association with solar flares pose particular challenges to theory. These are compounded for the long-lasting events observed by LAT \citep{2014ApJ...789...20A,2018ApJ...865L...7O,2018ApJ...869..182S}. Some role for a shock wave in the interplanetary medium seems likely. Radio observations offer some support for this idea \citep{2018ApJ...868L..19G} but questions persist, particularly around the feasibility of shock-accelerated particles making their way back to the Sun \cite[e.g.][]{2018IAUS..335...49H}. The absence of a strong correlation between the number of particles measured in space and the number deduced at the Sun \citep{2019ApJ...879...90D} is a further complication. By modelling the formation of the gamma-ray continuum in detail we may hope to constrain the angular and energy distribution of the primary ions and the properties of the medium where they interact, and thus to constrain particle acceleration models.

The FLUktuierende KAskade (FLUKA) code \citep{FLUKA,2014NDS...120..211B} is a general purpose Monte Carlo code for simulating the passage of energetic particles in matter, including the production and propagation of secondaries. It includes detailed, physically based, well validated  models for the directions and energies of secondaries produced in hadronic interactions. One defines one or more geometrical shapes and their chemical composition, and the spatial form, species and velocity distribution of the primary particles. FLUKA follows test particles and their secondaries throughout the whole of the defined volume(s). One may set ``detectors'' to carry out various tasks, e.g. determine the energy deposited in a particular volume, or the flux of particles of a particular type crossing the boundary between two regions. It was introduced as a tool for modelling solar flare gamma-radiation by \cite{2019SoPh..294..103T}. Here we will use FLUKA to predict the gamma-ray continuum radiation $\gtrsim 10$~MeV resulting from hypothesised energetic ion distributions interacting with the solar atmosphere and illustrate how these simulated spectra may be used to interpret observed gamma-ray spectra..

FLUKA is ``not a toolkit''. Its authors aim in the first instance to provide treatments of all the processes taking place, to the same degree of precision, rather than to offer the user a set of disconnected modules each providing its own treatment of a particular physical process. As a consequence it can be used to simulate a single, self-consistent gamma-ray spectrum extending from the positron annihilation line at 0.511 MeV through the nuclear deexcitation line component at $\approx 1 - 7$~MeV, to the pion decay continuum in the GeV energy range \citep{2019SoPh..294..103T}.  Nonetheless mechanisms are provided to, e.g., suppress some of the full set of processes taking place, even if this is strictly unphysical, and these prove very useful in developing physical understanding.   

In the next section we recall some basic kinematics of pion production in p-p collisions, highlighting the likely importance of the angular distribution for gamma-ray spectra. In Section~\ref{flintro} we make some comments on the use of FLUKA for astrophysical modelling and describe how we set up a model that can be used to simulate solar flare gamma-ray spectra. The core of Section~\ref{flgam} is a set of FLUKA-simulated thick target gamma-ray spectra for a range of assumed ion energy and angular distributions. Before this we  further illustrate some of the main physical processes. In Section~\ref{latflare} we compare some of the simulated spectra with Fermi LAT data of a solar flare. Section \ref{sconc} gives some concluding remarks. 

\section{Pion Production}
\label{basics}
\subsection{p-p Collisions}
\label{kinemat}

To see the potential importance of the ion angular distribution for the gamma-ray spectrum, we recall the threshold energy calculation for $\pi^0$ production in p-p collisions. Suppose an energetic proton with mass $m_p$ and velocity $v = \beta c$  collides with another proton at rest. The centre of momentum (CM) frame has
\begin{equation}
\gamma_{CM} = \sqrt{\frac{\gamma + 1}{2}}
\end{equation}
\noindent where $\gamma = (1-\beta^2)^{-1/2}$ and $\beta$ is the proton speed in units of the speed of light, as usual. The total CM energy is $E_{CM} = \sqrt{2\left(\gamma+1\right)}m_pc^2$. The proton energy at $\pi^0$ production threshold is determined by the requirement $E_{CM} = \left(2m_p +m_{\pi^0}\right)c^2$ and the $\pi^0$ is produced with $\gamma = 1.072$ \emph{in the laboratory (lab) frame}, moving in the same direction as the incident proton - not at rest, even although we consider the case of threshold. $\pi^0$ decay, isotropic in the CM frame, will take on some modest beaming along the incident proton direction when viewed in the lab frame. An observer in the backward direction will be more likely to see red-shifted photons and the gamma-ray spectrum will exhibit an asymmetry to lower photon energies. When it decays, the two emitted photons will in general be Doppler shifted, dependent on the direction of emission. In the extreme case that they are emitted parallel or anti-parallel to the pion lab frame velocity vector, their lab frame energies will be 43 MeV and 91 MeV.  

Above threshold more energy is available in the CM frame for sharing among the products of the reaction but there will still be a preference for the pions to be produced at a small angle to the incident proton. When primary ion distributions are isotropic, or if we average over the viewing angle, the $\pi^0$ decay spectrum is expected to be symmetric in $\log\epsilon$ about $m_{\pi^0}c^2/2 = 67$ MeV. In contrast, an ion distribution beamed away from the observer will produce a spectrum with an excess of red-shifted photons.  

Secondary electrons and positrons too will exhibit angular distributions that reflect the ion angular distribution. Their energy distribution displays a maximum at an energy of $\approx 35$~MeV \citep{1986ApJ...307...47D,1987ApJS...63..721M}, i.e. a Lorentz $\gamma \simeq 70$. Their bremsstrahlung radiation has a characteristic angular width of $\approx 1/\gamma \lesssim 1^{\circ}$ so again we can expect anisotropy of the emitting particles to have an important influence on the detected radiation. 

\begin{figure}   
   \centerline{\includegraphics[width=0.6\textwidth,clip=]{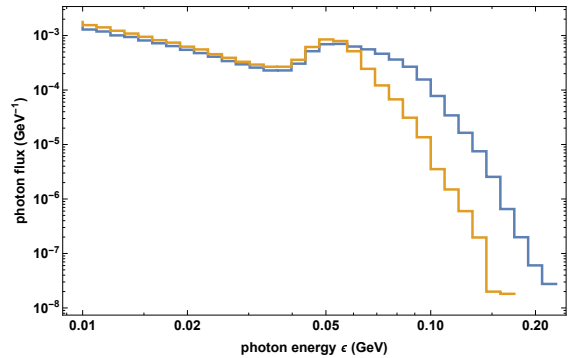}
              }
   \caption{Gamma-ray spectrum produced in the backward hemisphere from protons of initial energy 350 MeV directed downwards into a thick target source. The initial angular distribution is: unidirectional, parallel to the local vertical (yellow curve), or isotropic in the downward hemisphere (blue). The spectra are normalised to one primary proton.}
    \label{beamvsiso}
  \end{figure}

To further illustrate the effects of directionality, Figure~\ref{beamvsiso} shows the angle-averaged photon spectrum emitted into the backward hemisphere when 350 MeV protons are injected into a hydrogen target (more simulation details are discussed below, in Section~\ref{flintro}). Two different initial angular distributions are assumed: beamed, unidirectional downwards, and isotropic in the downward hemisphere. The resulting spectra have some notable features. While some e$^\pm$ bremsstrahlung continuum is produced, such low energy ions do not produce these secondaries prolifically and the $\pi^0$ component of the spectrum may be clearly distinguished, particularly in view of its narrowness. The peak of the $\pi^0$ spectrum is Doppler shifted from 67 MeV to more like 55 MeV, a  value consistent with lab frame $\pi^0$ energies as above. The peak is significantly sharper in the downward beamed case, with a factor of $\approx 5$ fewer photons above 60 MeV.  Effects of anisotropy are maximised just above threshold but we see in this example the potential of gamma-ray spectra to diagnose the primary ion angular distribution.

\subsection{Depth of Production}
\label{inside}

Ion energy distributions certainly extend on occasion into the GeV  energy range. A 1 GeV proton has a vertical stopping depth of 159 g~cm$^{-2}$ of hydrogen \citep{NISTBerger}, far beyond the $\approx 4$ g~cm$^{-2}$ depth of the photosphere \citep[e.g.][]{1981ApJS...45..635V}. Magnetic mirroring will prevent some ions from reaching such depths but any ions that can precipitate freely will emit pion decay secondary photons throughout most of this range. Considering 100 MeV photons for illustrative purposes, we find that the mean free paths against Compton scattering and pair production are 233 and 265 g~cm$^{-2}$, respectively \citep{NISTXCOM}. These are comparable to the depths from which many secondary photons originate so radiation will escape, but with a spectrum modified to some extent by these processes. Both processes imply the production of secondary, energetic electrons throughout some region of the atmosphere surrounding the field lines on which primary ions precipitate, whose bremsstrahlung will further modify the total photon spectrum escaping from the atmosphere.
  
\section{FLUKA Simulations} 
      \label{flintro}      

To set up a FLUKA simulation one prescribes the geometry, as a set of regions of various shapes, specifies the chemical composition of each region, fixes the parameters of the incident fast particle distributions and decides on a total number $N_{tot}$ of fast particles to be simulated. All of this information is encoded in an input file. 

For example, in most of Section~\ref{flgam} below we used a very simple, plane-parallel, two-region model of the solar atmosphere, divided in two by the plane $z=0$. $z$ measures distance downwards, into the Sun. In the region $z < 0$ we have a vacuum ``corona". In $z>0$ there is a uniform density ``atmosphere" of solar \citep{2009ARA&A..47..481A} composition. While a stratified density structure was implemented in \cite{2019SoPh..294..103T}, total thick target yields are insensitive to the detailed density structure so this simple model is adequate for our purposes here. The total depth of the atmosphere in $z$ and its lateral extent in $x$ and $y$ were made big enough to ensure all particles stop and produce their thick target yields within it.

As described in \cite{2019SoPh..294..103T} we wrote a custom subroutine to simulate particles coming from a power-law energy distribution between two fixed energies $E_{min}$ and $E_{max}$:
\begin{equation}
    N(E) \, = \, N_0 E^{-\delta} H(E-E_{min}) H(E_{max}-E) \, .
    \label{edist}
\end{equation}
\noindent Here $H$ is the Heaviside step function. The quantity $N_0$ takes care of normalisation; with the functional form fixed, FLUKA automatically normalises results to one primary particle. We generally take a value of $E_{min}$ determined by the threshold energy for the reactions of interest (e.g. 270 MeV for pion production in p-p collisions). $E_{max}$ and $\delta$ are parameters of the acceleration mechanism whose influence on the emergent photon spectra will be studied.

It seems likely that $E_{max}$ varies between events. At the upper extreme, the 26 February 1956 event produced signals at locations on Earth where the geomagnetic cutoff rigidity was $\approx 17$ GV \citep{2009AdSpR..44.1096R}. The ground level enhancement (GLE) event of 29 September 1989 evidently involved particles at Earth of up to 25 GV rigidity \citep{1990GeoRL..17.1073S} or possibly even higher \citep{2000SSRv...91..615M}. For the 24 May 1990 event, \cite{1997ApJ...479..997D} showed from energetic neutron timing that $E_{max} \approx 2$ GeV. \cite{1994SoPh..151..147A} argued that SMM/GRS spectra in the $> 10$ MeV range were best interpreted in terms of $E_{max}$ in the 0.5\textendash 0.8 GeV range. Thus the $E_{max}$ values found at the Sun could lie in a fairly wide range, from $\lesssim 1$ to $> 10$ GeV. To study its consequences here we parametrise this maximum energy via a sharp cutoff at $E_{max}$, rather than, e.g., an exponential roll-off so that any curvature in the simulated spectra is clearly due to radiation processes and transport, rather than the departure of the ion distribution from a power law.

We define $\mu = \cos\theta$, where $\theta$ is the angle between the ion velocity vector and the local (downward) vertical. Thus $\mu = 1$ corresponds to an ion directed vertically downwards and $\mu = 0$ to an ion that starts out parallel to the solar surface. The angular distribution of primary ions is described by the function $M(\mu)$. We implemented four possible forms: 
\begin{align}
M(\mu) & =  \delta_D(\mu-1) \nonumber \\
M(\mu) & =  H(\mu) \nonumber  \\
M(\mu) & =  2 \mu H(\mu) \nonumber \\
M(\mu) & =  \frac{3}{2} \left(1 - \mu^2\right) H(\mu) \, ,
\label{mudist}
\end{align}
\noindent i.e. unidirectional vertically downwards (we denote the Dirac delta function by $\delta_D$ to avoid confusion with the energy spectral index),  downward isotropic, moderately beamed downward, or a moderate ``pancake'' distribution, concentrated at large angles to the vertical. 

Once started, FLUKA generates $N_{tot}$ incident particles with energies drawn from the distribution given in Equation \ref{edist} and directions $\mu$ drawn from whichever form of $M(\mu)$ and follows them and their secondaries until they stop or, if they escape the dense atmosphere into the region $z<0$, reach the boundary of the box. 

Like all Monte Carlo codes, FLUKA can be made to output very large quantities of data describing details of the evolution of all particles. In this work, however, we followed the FLUKA authors' recommendation and implemented standard ``detectors", devices in the code that monitor all particles crossing the boundary between two regions, or evaluate the energy deposited in a particular volume by simulation particles. We set such a detector to detect photons passing through the boundary plane $z=0$ from the dense atmosphere to the corona. This gives us the spectrum of photons emitted into the backward hemisphere, rapidly and easily at the expense of some more detailed information about each photon. In the next few sections we integrate over all directions but it is also possible to study directionality by binning photons in solid angle, count statistics allowing. 

FLUKA carries out several repetitions of the specified Monte Carlo simulation to enable error estimation on derived quantities. We started the simulations described here by taking five repetitions (FLUKA default) with $N_{tot} = 6.4\times10^5$ each time, if necessary then increasing $N_{tot}$ and/or carrying out further repetitions so that the error in almost all photon energy bins was less than 3\% (except possibly where fluxes get very low at the highest energies).

\section{Simulated Gamma-Ray Spectra}
\label{flgam}

\subsection{Thin Target}

We first show some results for a thin target, a source which primary particles traverse with negligible energy degradation. We study initially isotropic ions released at the centre of a spherical target, aiming to eliminate features of the photon spectra that arise from particle anisotropy (cf. Section~\ref{kinemat}). Resulting photon spectra may be compared with previous calculations \citep{1987ApJS...63..721M,1994SoPh..151..147A,2003A&A...412..865V} and may also be relevant to ions contained in a low-density region.   

 We consider a sphere of hydrogen. Primary ions start from the origin of the sphere (${\bf r = 0}$), with an isotropic angular distribution and a specified energy distribution. The sphere has uniform density $n_H$ and a total radius $R$. Primary ions scatter negligibly in angle \citep[e.g.][]{1956pfig.book.....S} so every ion in the calculation will traverse a column depth negligibly different from $n_HR$. We integrate photons and other escaping secondaries over the surface of the sphere and include all escaping particles, so that the results are effectively integrated over angle to the sphere surface normal. 

\begin{figure}    
   \centerline{\includegraphics[width=0.6\textwidth,clip=]{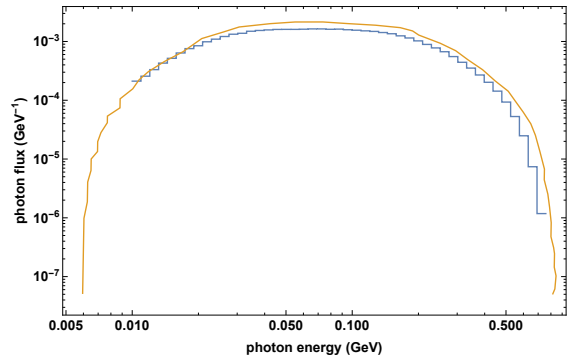}
              }
   \caption{FLUKA calculation of the gamma-ray spectrum produced by 1 GeV protons released isotropically at the centre of a sphere of hydrogen whose radius corresponds to a column depth of 0.084 g~cm$^{-2}$ (blue curve). For comparison the yellow curve shows the photon flux expected from the same situation calculated using the cross section given in \cite{1987ApJS...63..721M}. }
    \label{tpi0}
  \end{figure}

In Figure~\ref{tpi0} we show the photon spectrum escaping from such a source with $n_HR = 0.084$ g~cm$^{-2}$. All protons have an initial energy of 1 GeV. With this value of $n_HR$ they lose $\approx 4 \times 10^{-4}$ of their energy before leaving the target. The spectrum is dominated by $\pi^0$ decay photons with the expected form: symmetric in $\log\epsilon$, with a maximum at ~67 MeV \citep[e.g.][]{1970Ap&SS...6..377S}.  Also shown is the thin target flux expected from this situation calculated using the cross section for photon production via $\pi^0$ decay given in \cite{1987ApJS...63..721M}. The two independent calculations agree to within a factor of order unity, as we should expect in these spherically symmetric conditions. The result found using their results is slightly greater, by up to about 30\%, which we attribute to the different cross-section calculations of the two treatments. Similar agreement is obtained for other proton energies.

\begin{figure}    
   \centerline{\hspace*{0.015\textwidth}
               \includegraphics[width=0.515\textwidth,clip=]{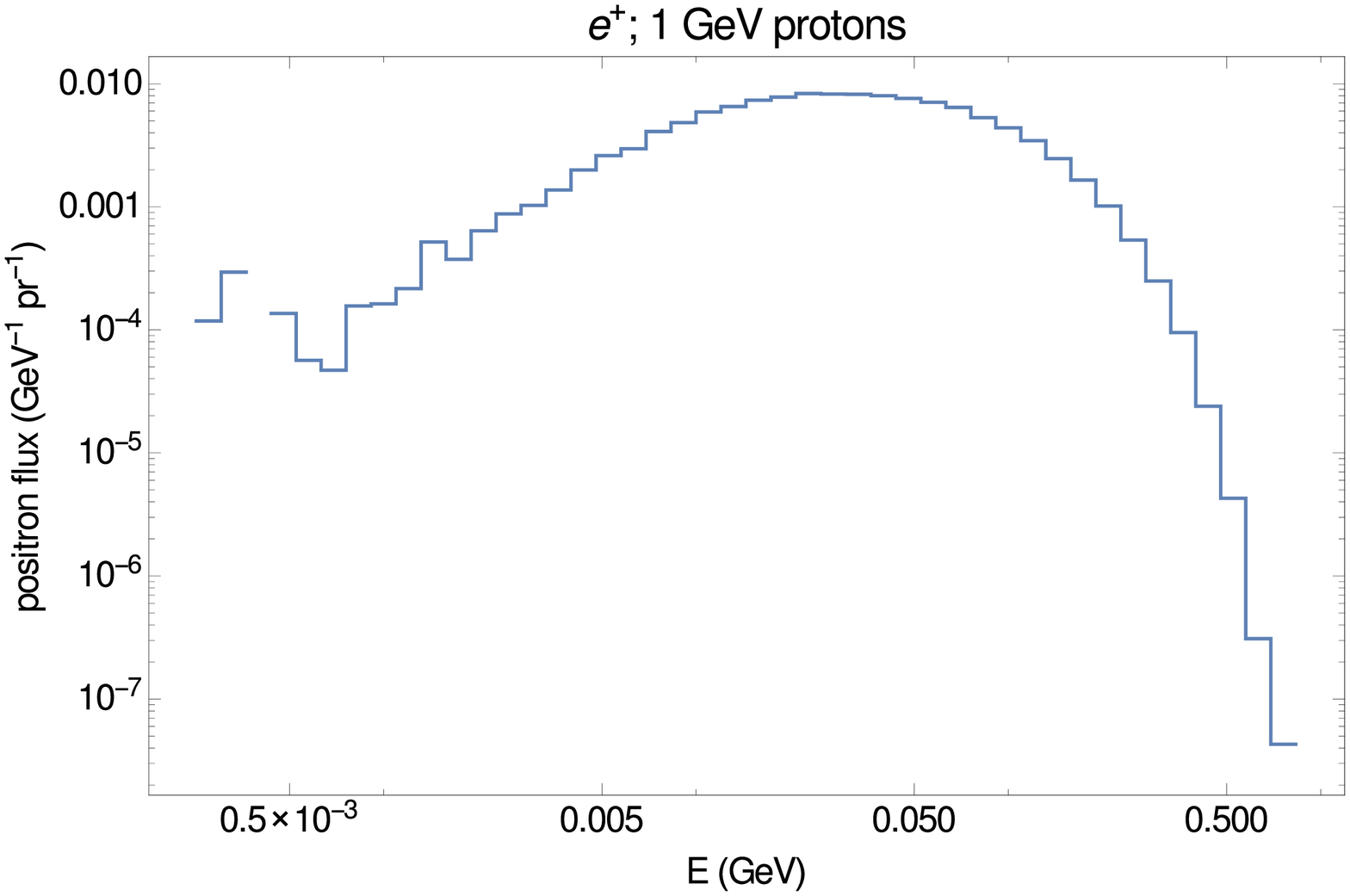}
               \hspace*{0.03\textwidth}
               \includegraphics[width=0.515\textwidth,clip=]{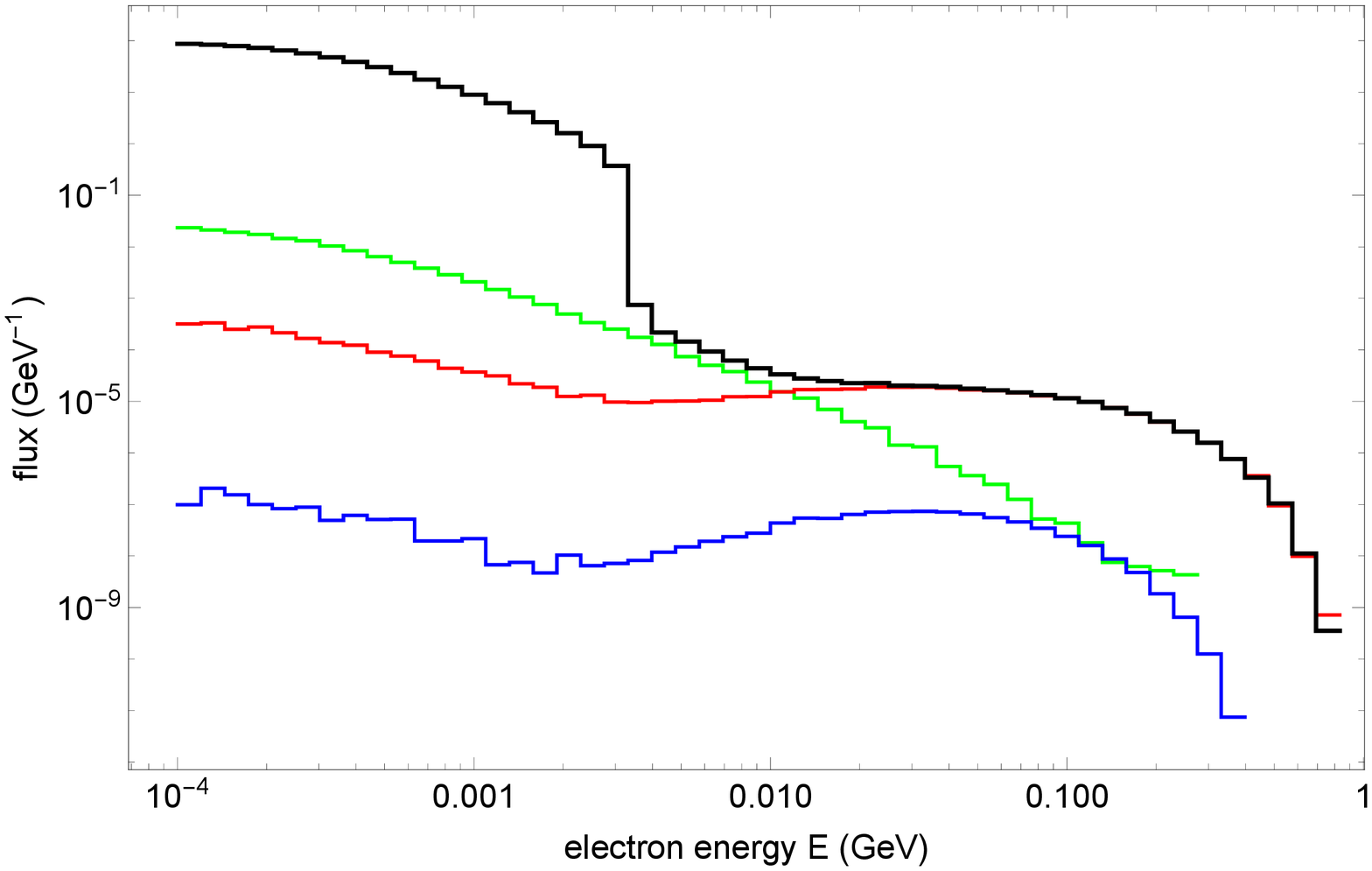}
              }
     \centerline{\Large \bf     
      \hspace{0.25 \textwidth}  (a)
      \hspace{0.415\textwidth}  (b)
         \hfill}

\caption{(a) Positron and (b) electron energy distributions leaving the surface of the spherical thin target source of Fig.~\ref{tpi0}. The distributions have been normalised to one primary proton. For electrons (panel b) we show the total electron distribution (thick, black curve), and the contributions via decay of: $\pi^-$ (blue), $\pi^0$ (red), $\pi^+$ (green).  
        }
   \label{thinepm}
   \end{figure}

The source is too thin for secondary e$^\pm$ to reveal themselves via bremsstrahlung but these particles are present nonetheless. Figure~\ref{thinepm} shows the electron and positron distributions that cross the boundary of the sphere. The positron distribution (Fig.~\ref{thinepm}a) has the form expected from $\pi^+ \rightarrow \mu^+ \rightarrow \mathrm{e}^+$ decay, roughly symmetric around a maximum at $\approx 35$ MeV \citep{1987ApJS...63..721M}. The sphere becomes a thick target for e$^{\pm}$ of around 1 MeV so the distribution at the lowest energies  has been modified by energy loss en route to the surface.

In Fig.~\ref{thinepm}b we show the total escaping electron distribution as well as the contributions produced via decay of each of $\pi^0$, $\pi^+$ and $\pi^-$ separately. The threshold energy for $\pi^-$ production is 870 MeV and the cross section is still very small at 1 GeV \citep{1986ApJ...307...47D}. Electrons are produced via $\pi^-$ decay only at about $10^{-6}$ of the level of positrons. Instead the Dalitz decay $\pi^0 \rightarrow \gamma \mathrm{e}^+ \mathrm{e}^-$, which occurs for $1.2\%$ of $\pi^0$ \citep{2014ChPhC..38i0001O}, accounts for most of the electrons above $\approx 10$ MeV. 

Below $\approx 3.3$ MeV the distribution is dominated by knock-on electrons, i.e. ambient electrons scattered to higher energies by primary protons. Consider an electron initially at rest, gaining energy at the expense of a proton moving initially at speed $\beta c$. Conserving energy and momentum \citep[e.g.][]{jackson,1966PhRv..150.1088A} we find the maximum energy $\Delta E$ gained by the electron to be
\begin{equation}
\frac{\Delta E}{m_ec^2} \, \approx \, 2\beta^2\gamma^2 \, ,
\label{kinemax}
\end{equation}    
\noindent where $\gamma = (1-\beta^2)^{-1/2}$ as usual. This expression is very close to the correct value as long as $\gamma \ll m_p/m_e$. For a 1 GeV primary proton Equation \ref{kinemax} gives a maximum knock-on electron energy of 3.3 MeV, as seen in Fig.~\ref{thinepm}b. The dominance of knock-on secondaries over pion decay products at low energies has already been discussed in the cosmic ray context by \cite{1966PhRv..150.1088A}. Implications of knock-on electrons for studying flare ions will be discussed elsewhere.

Even $\pi^+$ decay products can contribute to the \emph{electron} flux. Positrons will (Bhabha) scatter ambient electrons before leaving the sphere. Because positrons are so much more numerous this population may be comparable to electrons produced by $\pi^0$ or $\pi^-$ decay. We see from Fig.~\ref{thinepm}b that they are the dominant contribution to the flux leaving the sphere between 3.3 and $\approx 10$ MeV.

The relative magnitudes of the various components depend on specifics of this situation: the thickness of the sphere and the primary proton energy. In particular, once we consider protons further above the threshold for $\pi^-$ production we would expect secondary electrons via this channel to dominate those resulting from Bhabha scattering by secondary positrons. Nonetheless this example first reassures us that results from FLUKA are comparable to those obtained using other calculational approaches, and begins to illustrate the modifications introduced by transport in the source.

\subsection{Isotropic Thick Target}

\begin{figure}    
   \centerline{\includegraphics[width=0.6\textwidth,clip=]{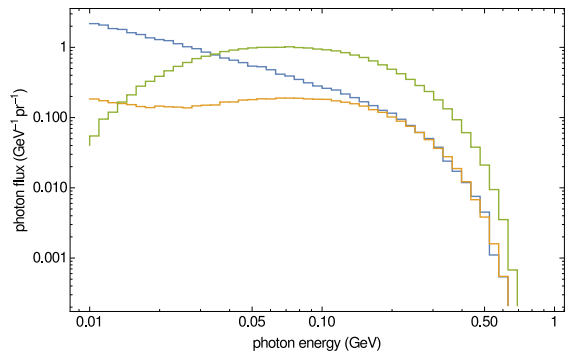}
              }
   \caption{Angle-integrated gamma-ray spectrum, per primary proton, produced in a spherical thick target source (blue curve). 1 GeV protons are released isotropically from the centre of a sphere of hydrogen with $n_HR = 160$ g cm$^{-2}$. The other two curves show the photon spectrum without the contribution from secondary e$^{\pm}$ (orange), and additionally with Compton scattering and pair production suppressed (green).}
    \label{thsphere}
  \end{figure}

With the aim of comparison with previous calculations, we next simulate a thick target source with the same spherical symmetry. Again we release an isotropic distribution of 1 GeV protons in a spherical target at ${\bf r = 0}$. Now $n_H$ and $R$ are chosen such that $n_HR = 160$ g~cm$^{-2}$, slightly greater than the stopping depth of a 1 GeV proton in hydrogen \citep[158.7 g~cm$^{-2}$, ][]{NISTBerger}. Again we show the angle-integrated energy distribution of photons crossing the boundary of the sphere, in Fig.~\ref{thsphere}.

Photons are present to $\approx 700$ MeV. The spectrum that emerges from the target steepens above 100 MeV but shows no pronounced $\pi^0$ "bump". For understanding we also show the spectra that result with certain physical processes artificially suppressed. First we set a very high threshold for the creation of secondary e$^{\pm}$, so that only $\pi^0$ decay photons are present. Now the spectrum below 100 MeV is flat, almost energy independent. We might expect such a spectrum if the source was threaded by a very high magnetic field so that e$^{\pm}$ lifetimes were suppressed. Lastly we also suppress Compton scattering and pair production, processes that modify the escaping spectrum. Only in this case do we reproduce the thick target $\pi^0$ spectra of previous authors \citep{1987ApJS...63..721M,1994SoPh..151..147A,2003A&A...412..865V}, with a maximum at $\approx 67$ MeV. 

We contrived this spherically symmetric situation for purposes of comparison with previous calculations. The primaries start out at the maximum possible column density, a situation we do not expect to occur on the Sun. Nonetheless we can conclude that FLUKA gives results in agreement with previous calculations, and that protons in the GeV energy range may produce observable radiation in spite of their great stopping depths, but with a spectrum significantly modified by pair production and Compton scattering.

\subsection{Downward Directed Primary Ions: Power Law}
Here we show a set of simulated backward hemisphere gamma-ray spectra from primary protons injected with power-law energy distributions and the angular distributions specified in Equations~\ref{mudist} above. We concentrate on the continuum spectrum, above 10 MeV; line features at lower energies were discussed by \cite{2019SoPh..294..103T}. For now spectra are averaged over the viewing angle but we discuss directionality in the next section.

  \begin{figure}    
   \centerline{\hspace*{0.015\textwidth}
               \includegraphics[width=0.49\textwidth,clip=]{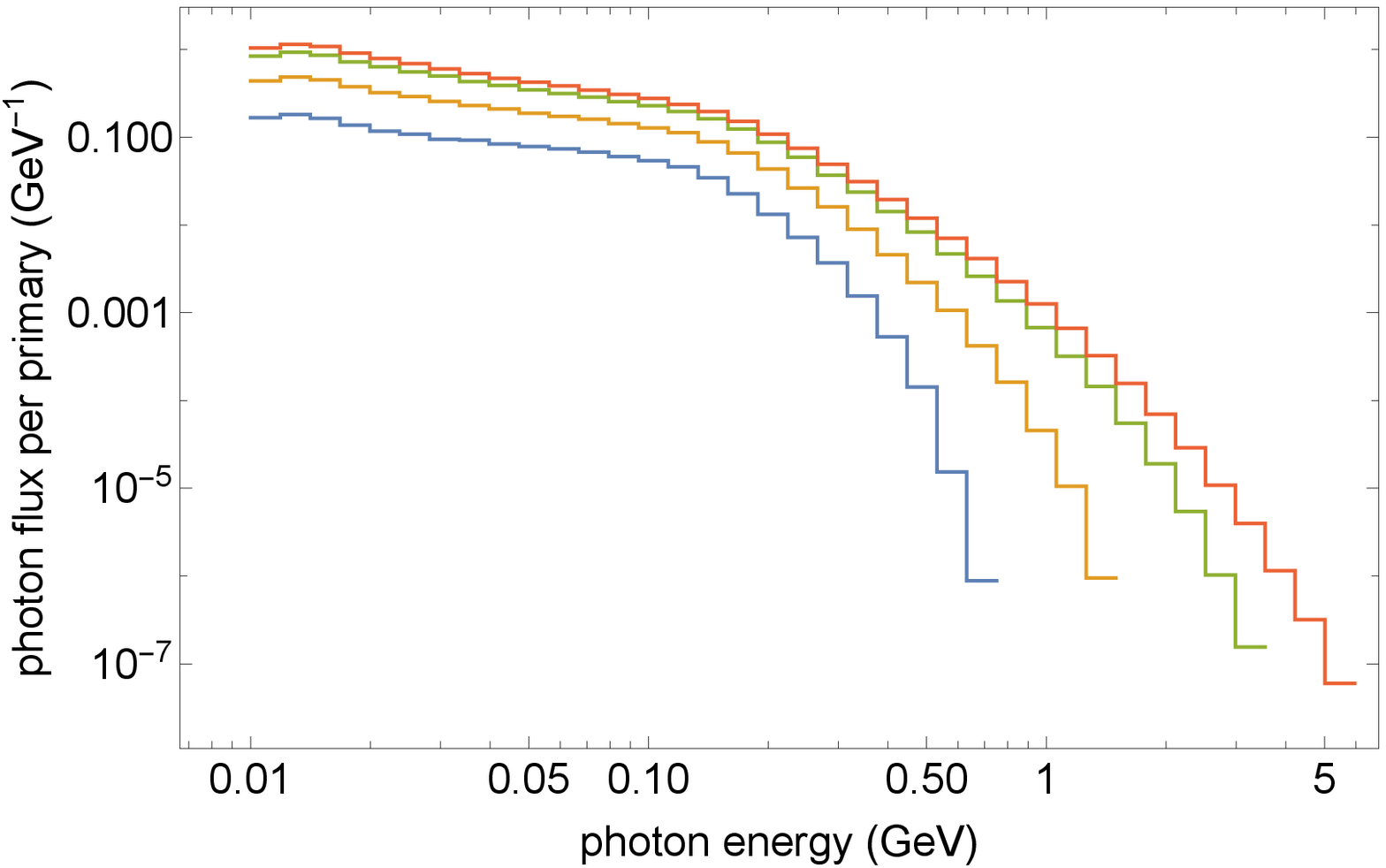}
               \hspace*{0.06\textwidth}
               \includegraphics[width=0.49\textwidth,clip=]{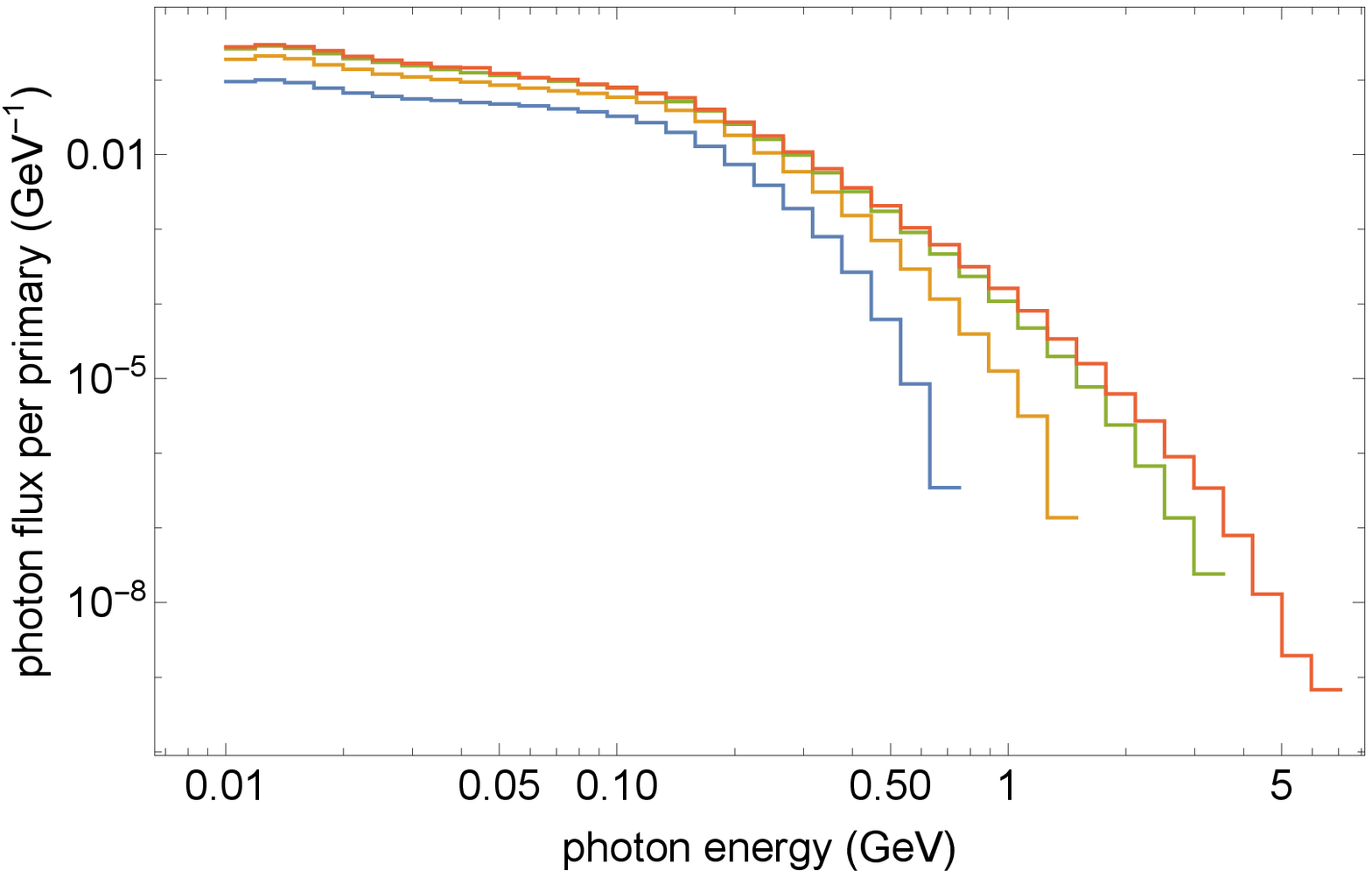}
              }
    \vspace{-0.30\textwidth}   
     \centerline{\Large \bf     
      \hspace{0.4\textwidth}  \color{black}{(a)}
      \hspace{0.48\textwidth}  \color{black}{(b)} 
         \hfill}
     \vspace{0.31\textwidth}    
           
   \centerline{\hspace*{0.015\textwidth}
               \includegraphics[width=0.49\textwidth,clip=]{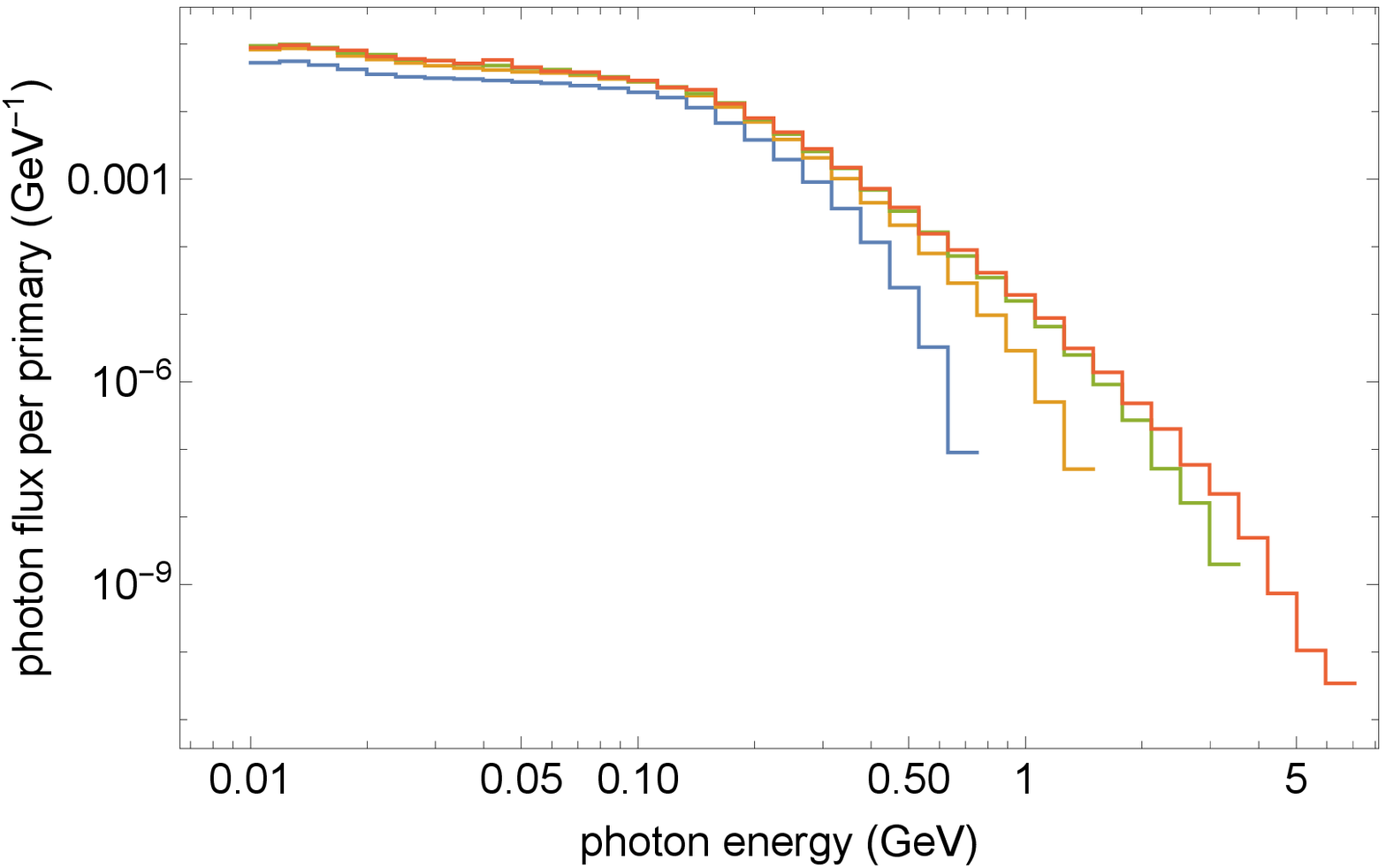}
               \hspace*{0.06\textwidth}
               \includegraphics[width=0.49\textwidth,clip=]{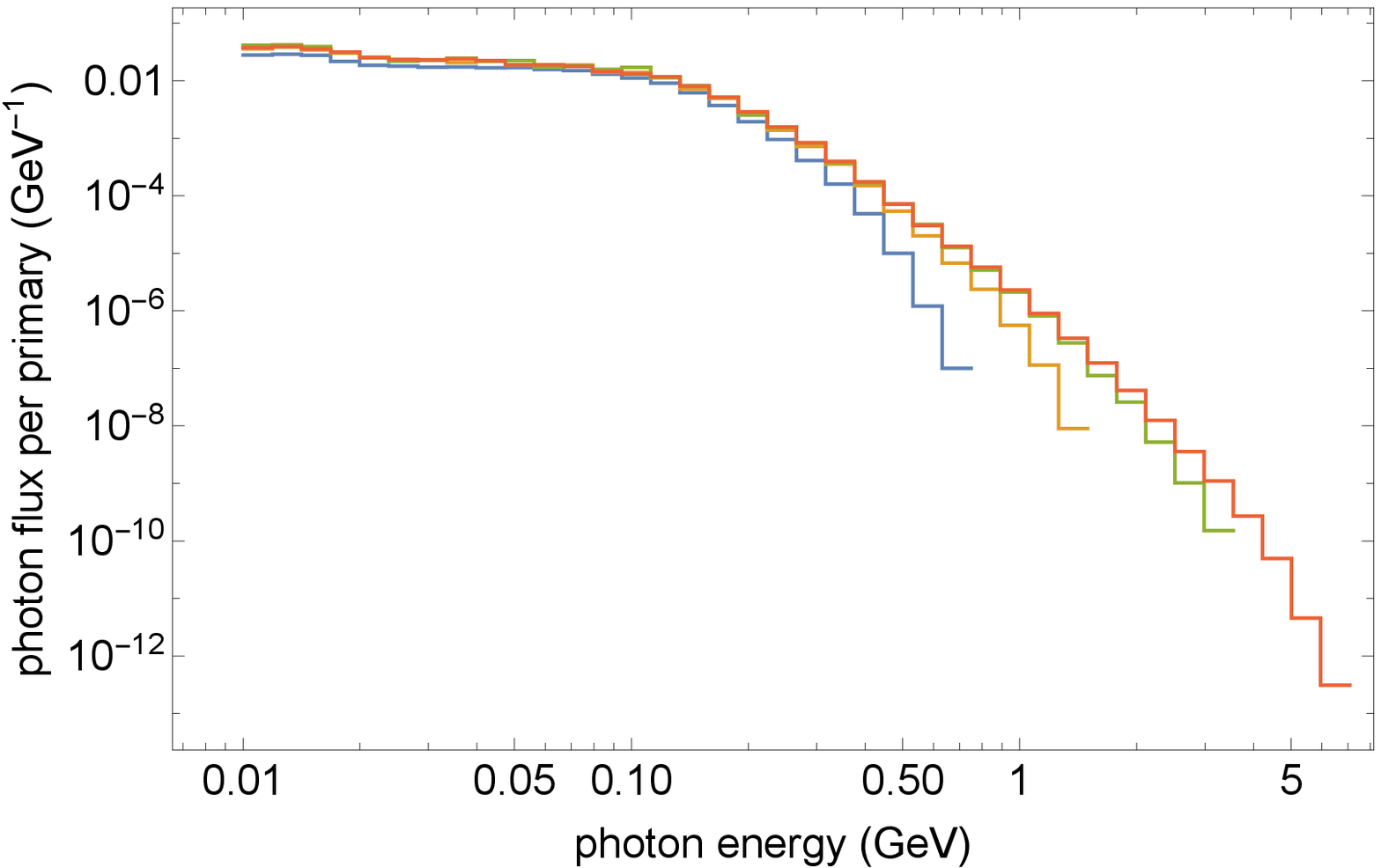}
              }
     \vspace{-0.3\textwidth}   
     \centerline{\Large \bf     
      \hspace{0.4\textwidth}  \color{black}{(c)}
      \hspace{0.48\textwidth}  \color{black}{(d)}
         \hfill}
     \vspace{0.31\textwidth}    
              
\caption{Gamma-ray spectra from downward isotropic protons impinging on a target of solar composition. Fluxes are normalised to one proton between 0.27 MeV and $E_{max}$. Energy distributions have $\delta = 2$ (panel a), 3 (panel b), 4 (panel c) and 5 (panel d). In each figure, $E_{max} = 1$ (blue), 2 (gold), 5 (green) and 8 GeV (red). 
        }
   \label{isospect} 
   \end{figure}

 \begin{figure}   
   \centerline{\includegraphics[width=0.6\textwidth,clip=]{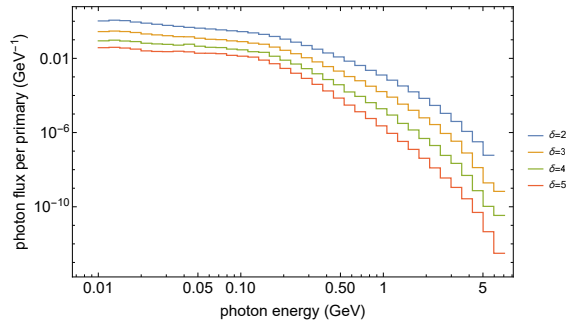}
              }
   \caption{Gamma-ray spectrum produced in the backward hemisphere from power-law distributed protons between 0.27 MeV and 8 GeV, with $\delta = 2, 3, 4, 5$, isotropic in the downward hemisphere.   The spectra are normalised to one primary proton.}
    \label{e8d2345}
  \end{figure}

We use the same geometry as above and also now include He, C, N, and O in the composition of the target, with the relative abundances of \cite{2009ARA&A..47..481A}.  Figure~\ref{isospect} shows photon spectra resulting from primary protons injected into this source with a downward isotropic distribution. Each panel is for a fixed value of $\delta$: 2, 3, 4 or 5 (panels a, b, c, and d respectively). Within each panel we show four spectra, calculated using different values of $E_{max} = 1$, 2, 5 or 8 GeV. Higher-energy protons have a greater probability to produce pions so the net photon yield per primary increases with $E_{max}$, and decreases with $\delta$. As $E_{max}$ increases, the photon spectrum extends to higher photon energies and the spectrum above $\approx100$ MeV exhibits more power-law character.

With $E_{max} = 1$~GeV the form of the photon spectrum barely depends on $\delta$: proton energies do not extend far enough above threshold for the spectrum to bear an imprint of the proton energy distribution. In Figure~\ref{e8d2345} we bring together the four spectra calculated with $E_{max} = 8$~GeV to show more clearly the steepening of the photon spectrum with increasing $\delta$. The spectrum has an approximate power-law character $\approx \epsilon^{-\eta}$ above about 100 MeV, steepening more rapidly as the upper cutoff is approached. In the photon energy range 0.1 \textendash 2 GeV, with this value of $E_{max}$ an increase of 1 in $\delta$ leads to an increase of $\approx0.5$ in $\eta$.

 \begin{figure}   
   \centerline{\includegraphics[width=0.6\textwidth,clip=]{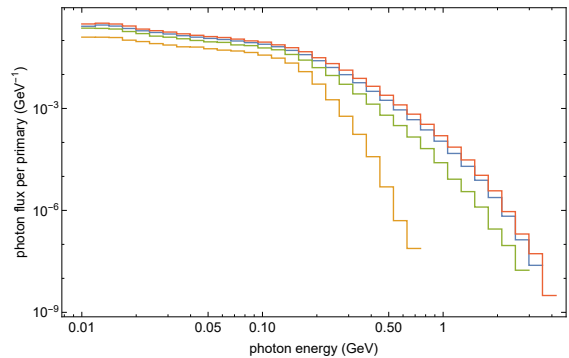}
              }
   \caption{Gamma-ray spectra produced in the backward hemisphere from protons with $\delta=3$, $E_{max} = 5$~GeV and the four angular distributions of Equations \ref{mudist}: pancake distribution (red), downward isotropic (blue), proportional to $\mu$ (green), and unidirectional vertically down (gold).}
    \label{d3e5ang}
  \end{figure}

\begin{figure}   
   \centerline{\includegraphics[width=0.6\textwidth,clip=]{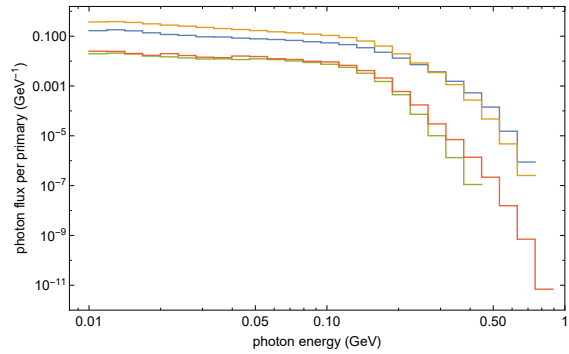}
              }
   \caption{Gamma-ray spectra produced in the backward hemisphere from protons with a unidirectional, vertically downward angular distribution, with $\delta=2$ and $E_{max} = 1$~GeV (blue), $\delta=2$ and $E_{max} = 8$~GeV (gold), $\delta=5$ and $E_{max} = 1$~GeV (green), $\delta=5$ and $E_{max} = 8$~GeV (red).}
    \label{unid}
  \end{figure}

In Figure \ref{d3e5ang} we fix $\delta = 3$ and $E_{max} = 5$~GeV and calculate emergent photon spectra for each of the four primary angular distributions in Equations~\ref{mudist}. As expected we see that a greater proportion of downward-directed primaries leads to fewer escaping photons, with a spectrum falling off more steeply with photon energy. In the limiting case $M(\mu) = \delta_D(\mu-1)$ the escaping spectrum falls off very steeply and cuts off at a much lower energy than the other cases. In this extremely beamed case the escaping photon spectrum is almost completely insensitive to the details of the primary proton distribution, never extending above  $\approx700$~MeV. We illustrate this further in Fig.~\ref{unid}, showing results for this form of $M(\mu)$ with $E_{max} = 1$ or 8 GeV, and $\delta = 2$ or 5. While the cases with $\delta = 2$ produce a greater photon flux and the cases with $E_{max} = 8$~GeV extend to slightly higher photon energies, it would be difficult observationally to distinguish between these spectra, despite the very different ion distributions that produce them. 

\subsection{Photon Spectrum Dependence on Viewing Angle}

The FLUKA detectors measure fluxes of particular particles species at boundaries or within defined volumes. We used the detector called USRBDX to obtain photon distributions averaged over the whole of the backward hemisphere, shown in the previous section, and also broken down into intervals of solid angle referred to the normal to the boundary (in our case, between the dense atmosphere and the corona). By default we divide the backward hemisphere, $0 < \theta < 90\degree$ into ten equal intervals of solid angle. As a first example, Fig.~\ref{d3e5ang} shows photon spectra calculated assuming an isotropic downward primary proton distribution with $\delta = 3$ and $E_{max} = 5$ and 8~GeV, averaged over viewing angles $\theta$ of $0\degree - 26\degree$ and $84\degree - 90\degree$. If the magnetic field is vertical these values correspond to a flare at disc centre and at the limb, respectively and we use these words to refer to the two cases although the true situation may be more complex. We see pronounced differences between the two cases: the disc centre spectrum exhibits no photons above $\approx 1$~GeV and no influence of $E_{max}$, similarly to the unidirectional downward cases of Fig.~\ref{unid}, whereas the limb spectra exhibit power-law tails extending to $\approx 4$ and 7~GeV. However the spectral flattening below $\approx 100$ MeV expected from $\pi^0$ decay is more pronounced for the disc centre spectra. 

\begin{figure}   
   \centerline{\includegraphics[width=0.6\textwidth,clip=]{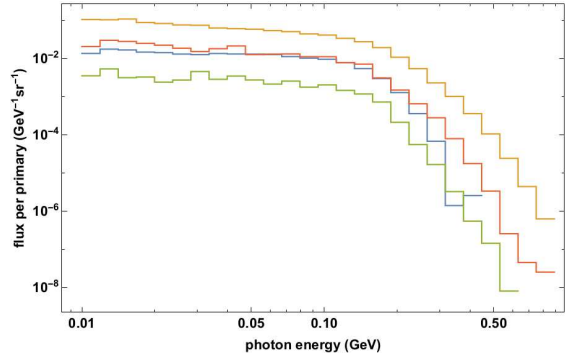}
              }
   \caption{Gamma-ray spectra emitted by flares at disc centre with: $\delta = 2$, $E_{max} = 1$~GeV and $M(\mu)=1-\mu^2$ (blue); $\delta=2$, $E_{max} = 8$~GeV and $M(\mu)$ isotropic (gold); $\delta = 5$, $E_{max} = 2$~GeV and $M(\mu)$ isotropic (green); $\delta=4$, $E_{max} = 8$~GeV and $M(\mu)$ isotropic (gold) (red). }
    \label{varydisk}
  \end{figure}

Disc centre spectra are dominated by the directionality of secondaries, together with transport of photons produced deeper in the atmosphere. Figure~\ref{varydisk} emphasises this, showing disc centre spectra with a variety of primary proton distributions. Only the normalisation of the photon spectrum varies significantly across these various cases and it would be difficult or impossible to discriminate between them observationally, even although $\delta$, $E_{max}$, and $M(\mu)$ are quite different. As a random example, note that the photon spectrum produced when  $\delta=2$, $E_{max}=1$~Gev and a pancake angular distribution is almost identical, up to about 200 MeV photon energy, with the photon spectrum produced with $\delta=4$, $E_{max}=8$~Gev and a downward isotropic distribution.

\begin{figure}   
   \centerline{\includegraphics[width=0.6\textwidth,clip=]{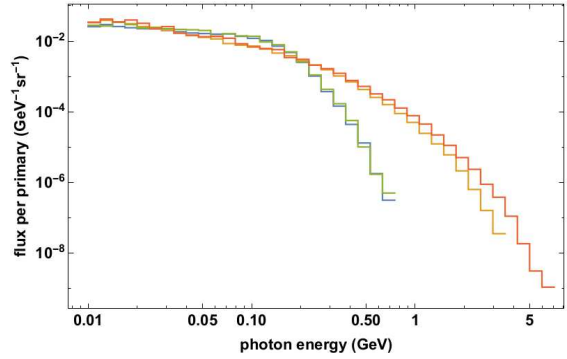}
              }
   \caption{Gamma-ray spectra produced from protons with a downward isotropic angular distribution and $\delta=3$, for flares at disc centre with $E_{max} = 5$~GeV (blue) and 8~GeV (green), and at the limb with $E_{max} = 5$~GeV (gold) and 8~GeV (red). }
    \label{d3e58}
  \end{figure}
  
\begin{figure}   
   \centerline{\includegraphics[width=0.6\textwidth,clip=]{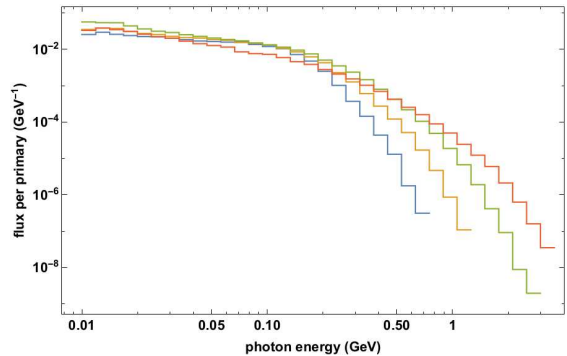}
              }
   \caption{Gamma-ray spectra emitted by flares assuming primary ions with $\delta = 3$, $E_{max} = 5$~GeV and a downward isotropic angular distribution, at different positions on the solar disc. Heliocentric angles are: $0\degree - 26\degree$ (blue), $46\degree - 54\degree$ (gold), $66\degree - 73\degree$ (green) and $84\degree - 90\degree$ (red), representing equal intervals of solid angle.  }
    \label{viewang}
  \end{figure}

At disc centre the form of the photon spectrum is very insensitive to the form of the primary proton distribution. At the limb the photon spectrum extends into the GeV energy range, increasing with $E_{max}$ and with hardness determined by the value of $\delta$.  As the viewing angle increases from $0\degree$ (disk centre) to $90\degree$ (limb) there is a smooth transition between these two sorts of spectrum, as shown in Fig.~\ref{viewang}. 

\section{Flare of 10 June 2010}
\label{latflare}

The Geostationary Operational Environmental Satellite (GOES) M2 class flare of 12 June 2010 (00:50 UT) was observed by the Large Area Telescope (LAT) of the NASA Fermi satellite \citep{2012ApJ...745..144A}. High count rates in the anti-coincidence shields meant that the flare was not detected using the standard LAT data analysis method but it was possible to obtain useful data using the LAT Low Energy (LLE) procedure \citep[detailed in the Appendix of ][]{2014ApJ...789...20A}. To show the usefulness of FLUKA for analysing flare data we obtain fits to the LLE spectrum from this flare. Photons were observed from this flare up to a maximum energy of $\approx 300$~MeV. The H$\alpha$ flare took place in an active region at N$23\degree$W$43\degree$, i.e. a heliocentric angle of $\approx48\degree$. 

 Photons consistent with a solar origin were detected in the LAT for about 60s, a much short duration than many other, highly extended events observed by LAT \citep[e.g.][]{2014ApJ...789...20A,2018ApJ...869..182S}. In the extended events it seems likely that ions will be trapped in low density regions by either or both of magnetic inhomogeneity (mirroring) and plasma turbulence, physical processes not included in our FLUKA simulations. Even in the brief 12 June 2010 event these effects could still be important. For example, a 1 GeV proton mirroring at the 4 g cm$^{-2}$ depth of the photosphere needs $\approx 40$ traversals of the loop to stop, a time of about 1.5~s for a $10^9$~cm loop length (neglecting any pitch-angle scattering) - much less than the event duration. However this short duration also means that radiation from ions trapped in the corona will be unimportant for the observed spectrum which is why we select this particular event for a first trial of FLUKA spectra against data. The emergent gamma-ray spectrum is determined by the energies and directions of the fast ions as they precipitate into the dense atmosphere, however complex their histories in the corona may have been. 
 
 \cite{2019SoPh..294..103T} previously showed that a FLUKA angle-averaged spectrum can provide an acceptable fit to these data, giving results similar to those found using the templates included in the standard data analysis software Object SPectral EXecutive (OSPEX) \citep{ospex}. Here we show additionally that the LAT data can be used to discriminate between different assumptions for primary ion energy and angle distributions. We neglect the contribution from primary $\alpha$-particles, straightforward to include but yielding little extra insight for present purposes. 

If magnetic field lines were all oriented vertical to the local solar surface, the viewing angle would be simply determined by the heliocentric angle but the real situation is likely to be more complex. \cite{2003ApJ...595L..81S} found that red shifts of nuclear deexcitation lines in the flare of 23 July 2002 were greater than expected on the basis of its heliocentric angle. \cite{2020MNRAS.491.4852H}, using the Bifrost MHD model, found that field lines may meander substantially in the chromosphere, presenting a wide range of angles to any particular line of sight. Thus we also allow the viewing angle to vary, rather than fixing it based on the heliocentric angle.

\begin{figure}
   \centerline{\hspace*{0.015\textwidth}
               \includegraphics[width=0.515\textwidth,clip=]{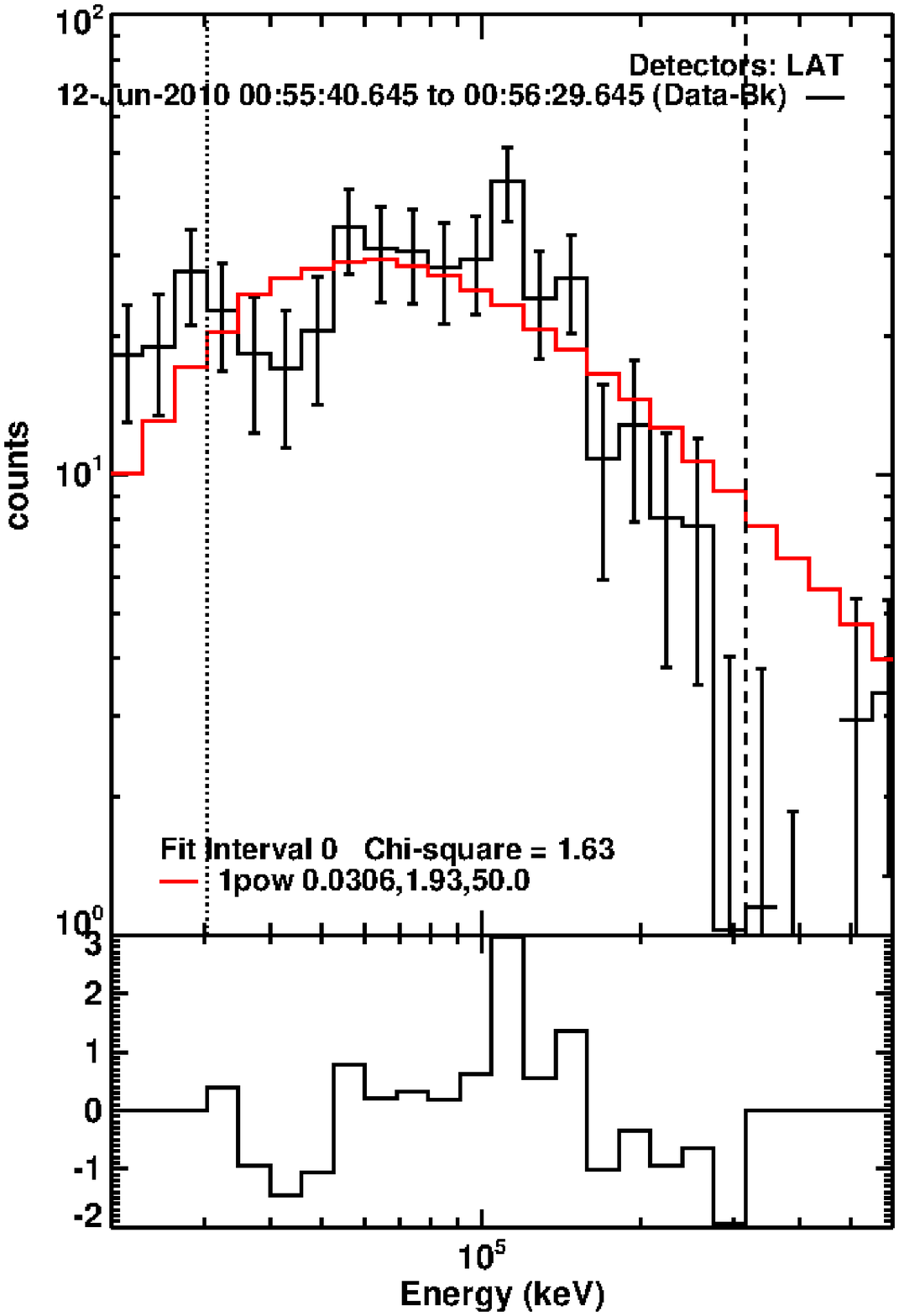}
               \hspace*{-0.03\textwidth}
               \includegraphics[width=0.515\textwidth,clip=]{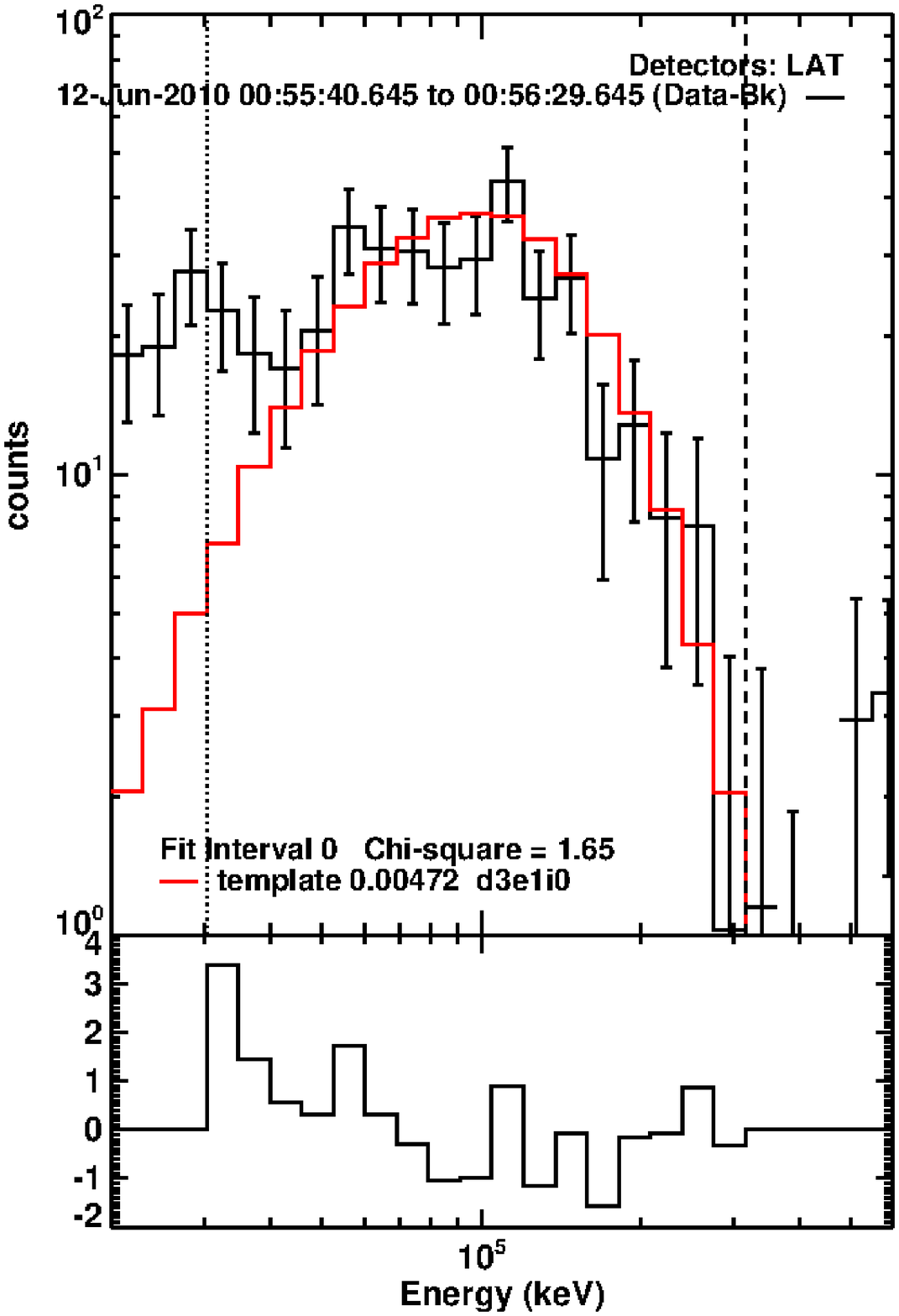}
              }

\caption{LAT count spectra from the 12 June 2010 flare together with the best-fit single power law (left) and a single pion decay template (right), calculated assuming a downward isotropic proton distribution with $\delta=3$, $E_{max} = 1$~GeV, and a viewing angle of $0\degree-25.8\degree$ to the downward vertical. The vertical broken lines show the photon energy range, $30 - 300$~MeV, in which the spectral fitting was carried out. In each case the bottom panel shows the normalised residuals for the fit. 
        }
   \label{fit1}
   \end{figure}

Following \cite{2012ApJ...745..144A} we fit the spectrum in the photon energy range 30 - 300 MeV, having subtracted an instrumental background calculated by interpolation between time intervals before and after the flare. We start by assuming a single power-law photon spectrum $I(\epsilon) = A\epsilon^{-s}$ and find a best-fit spectral index $s = 1.93$, in good agreement with that found by \cite{2012ApJ...745..144A} (Figure~\ref{fit1}). We also show an example of a fit to a spectrum including only a FLUKA pion decay template, calculated assuming $\delta=3$, $E_{max} = 1$~GeV, downward isotropic primary ions and a viewing angle (i.e. angle between the line of sight and the downward vertical) in the range $45.6\degree - 53.1\degree$, appropriate to the heliocentric angle of the flare. Only the amplitude of the spectrum, equivalent to the total number of protons, is determined by fitting the LAT data. The normalised $\chi^2$ values from these two fits are comparable, 1.63 (power-law) vs. 1.65 (pion decay). In both cases the pattern of residuals highlights systematic discrepancies between the model spectrum and the data: the observed spectrum has a curvature inconsistent with a single power-law; and the steeply falling spectrum in the $\sim 30 - 60$ MeV photon energy range is inconsistent with any of the pion decay templates. Thus we next fit the LAT data using the sum of a pion decay template and a single power law. In each case the parameters of the primary proton distribution are fixed and we find the best-fit values of three parameters: the amplitude of the pion-decay template (i.e. the total number of protons) and the parameters $A$ and $s$ of the power-law component.

\begin{figure}   
                \centerline{\hspace*{0.015\textwidth}
               \includegraphics[width=0.515\textwidth,clip=]{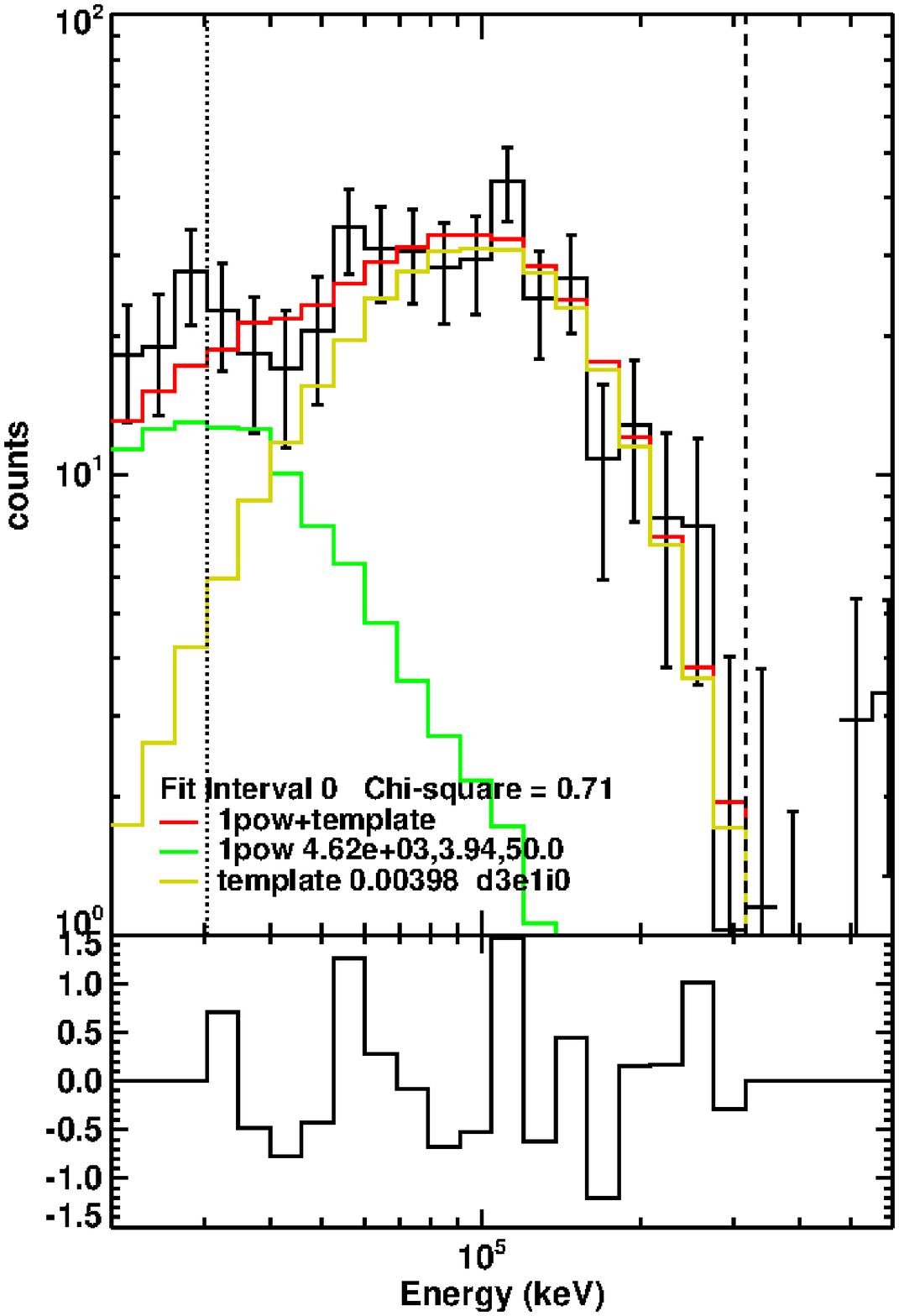}
               \hspace*{-0.03\textwidth}
                \includegraphics[width=0.515\textwidth,clip=]{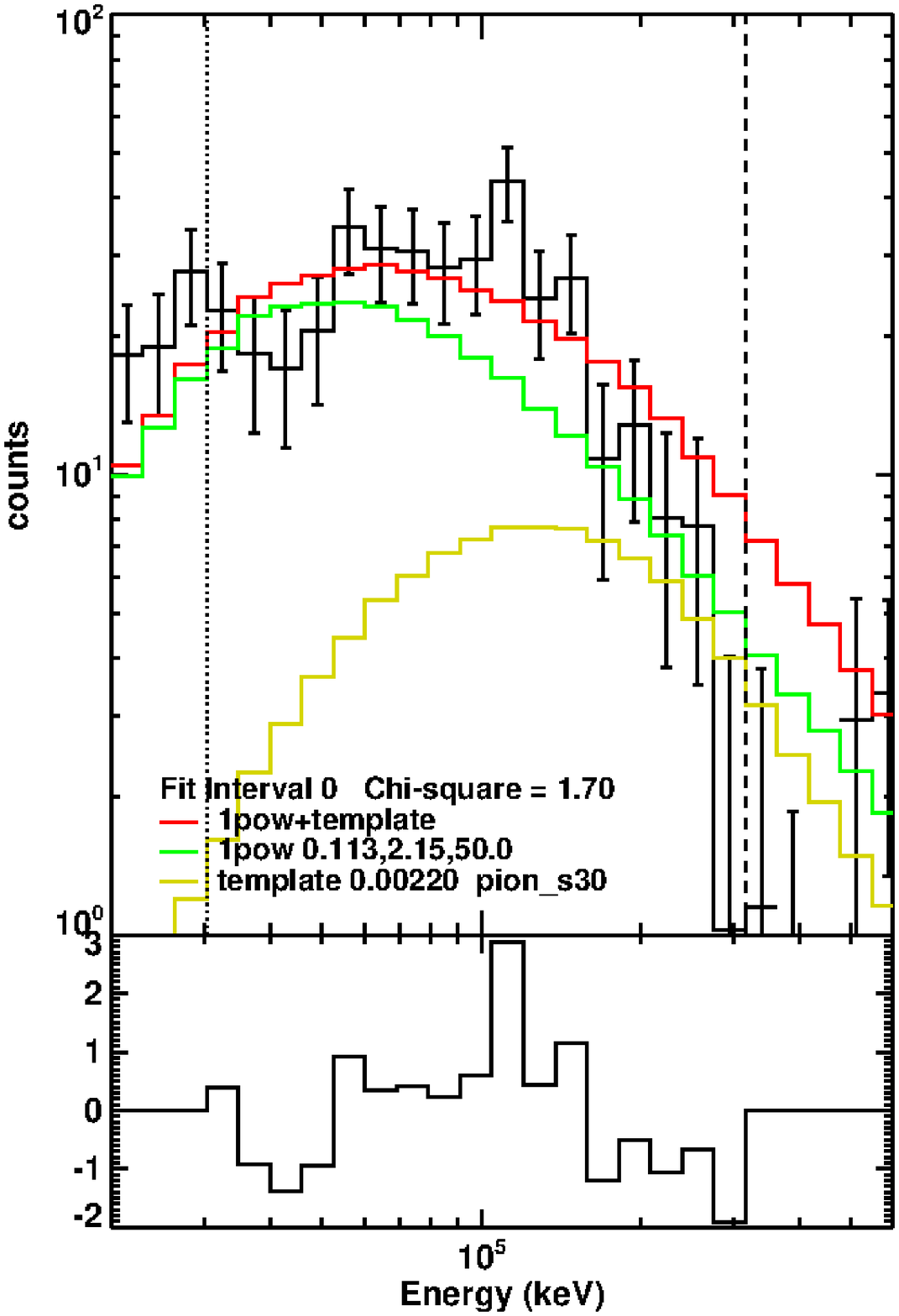}
              }
   \caption{Left: LAT count spectrum from the 12 June 2010 flare (as in Figs.~\ref{fit1}) together with the closest fit ($\chi^2 = 0.71$) found for a combination of a single power-law and a FLUKA template. The power-law spectrum has $s=3.94$ and the primary protons have $\delta=3$, $E_{max}=1$~GeV, and a downward isotropic angular distribution. The viewing angle is in the range $0\degree-25.8\degree$ to the downward vertical. The lower panel shows the residuals. Right: the same data together with the best fit obtained using a single power-law photon spectrum plus the OSPEX-supplied template for $\delta=3$.  }
    \label{bestfit}
  \end{figure}

We found best fits to the LAT spectrum for a range of primary ion parameters. Results are given in Table~\ref{params}. In Figure~\ref{bestfit} we show the fit that gave the smallest $\chi^2$ (0.71) from all those we tried, obtained for $\delta=3$, $E_{max} = 1$~GeV, downward isotropic ions and a viewing angle of $0\degree - 25.8\degree$. The power-law component is important in the 30-60 MeV energy range but the pion decay template dominates above this; modulating the power-law component with an exponential rollover $e^{-\epsilon/\epsilon_{roll}}$ does not improve the fit, whether the rollover energy $\epsilon_{roll}$ is fixed e.g. at 100 MeV or allowed to vary freely along with the other parameters. Also shown in Figure~\ref{bestfit} is the best fit obtained using the pion decay template supplied in OSPEX for $\delta=3$ together with a power-law photon spectrum. The template also assumes a magnetic field strength of 300 G (determining the $e^\pm$ synchrotron loss rate), and ambient density of $10^{15}$ cm$^{-3}$. Based on a homogeneous source with an isotropic distribution of ions, the OSPEX template resembles the FLUKA angle-averaged spectra given above (Figure~\ref{isospect}). As measured by $\chi^2$ the fit is poorer than the best-fit FLUKA spectrum (1.70 vs.  0.71), showing the importance of retaining particularly the angular dependence of the emission. The OSPEX template fit implies just $4.1\times10^{27}$ protons $>30$~MeV, three orders of magnitude less than found with the FLUKA fit, as a result of the greater weight given by OSPEX to the power-law spectral component. As shown also in the distribution of residuals, the angle-averaged OSPEX template has less of the curvature needed to play a substantial role in fitting the observed spectrum; cf. the fits shown in \cite{2019SoPh..294..103T}.

Many other primary proton parameters give fits that are equally acceptable, statistically, but some sets of parameters are clearly disfavoured and $\chi^2$ values show informative systematic trends. From Table~\ref{params} we can make some general statements:

\begin{itemize}

\item The lowest $\chi^2$ values are obtained with $\delta = 3, 4$. Assuming $\delta = 2$ or 5 leads to $\chi^2$ significantly greater than one, irrespective of other assumptions made 
\item Assume that primary protons are injected into the source with an isotropic distribution \citep[as might be expected for many particle acceleration mechanisms - e.g.][]{1974SoPh...37..353M}. Keeping $\delta$ and $E_{max}$ fixed, the best fits (i.e. lowest $\chi^2$) are then obtained for viewing angles close to $0\degree$, rather than the flare heliocentric angle. $\chi^2$ increases monotonically with viewing angle. Also $E_{max}$ should be comparatively low, 1 or 2 GeV (as in Fig.~\ref{bestfit}).
\item If we now suppose that ions are injected either unidirectionally downwards, or more modestly beamed $\sim \mu$, the lowest $\chi^2$ values with fixed $\delta$ and $E_{max}$ are generally found in the region of the flare location. $\chi^2 < 1$ may now be obtained with $E_{max} = 5$ or 8 GeV.
\item Viewing angles of $> 70\degree$ are always less consistent with the LAT data: $\chi^2$ values are greater and the power-law component plays a greater role throughout the photon energy range being fit.
\end{itemize}

\begin{table} 
\label{params}
\caption{Parameters for various assumed primary ion distributions and viewing angles, as found from fitting the 21 June 2010 LAT spectrum in OSPEX. $N_p$ is the number of protons above 30 MeV assuming the power law continues smoothly to that energy. Ions are assumed downward isotropic except when the spectral parameters $E_{max}$ and $\delta$ are marked with: an asterisk (*), in which case ions are unidirectional vertically downward; or a dagger $\dagger$, in which case they are moderately beamed, $\sim \cos\theta$. The top line gives the best fit single power law, with no contribution from ions, and a few lines give $N_p$ and $\chi^2$ when no power-law component is included. }
\begin{tabular}{ |c|c|c|c|c|c|c| }
\hline 
$ s $ & $A$ & $\delta$ & $E_{max}$ (GeV) & $N_p(> 30 \mathrm{MeV})$ & $\theta_{obs}$ & $\chi^2$ \\
\hline
1.93 & 0.0306 & - & - & - & - & 1.93 \\
\hline
3.09 & 29.3 & 2 & 2 & $9.5\times10^{28}$ & $0\degree - 25.8\degree$ & 0.85 \\
2.44 & 0.635 & 2 & 8 & $2.6\times10^{28}$ & $0\degree - 25.8\degree$ & 1.41 \\
- & - & 2 & 8 & $5.3\times10^{28}$ & $0\degree - 25.8\degree$ & 2.61 \\
1.94 & 0.0328 & 2 & 8 & (negligible) & $84.3\degree - 90\degree$ & 1.75 \\
3.94 & 4620.0 & 3 & 1 & $4.1\times10^{30}$ & $0\degree - 25.8\degree$ & 0.71 \\
- & - & 3 & 1 &  $4.9\times10^{30}$ & $0\degree - 25.8\degree$ & 1.65 \\
2.09 & 0.0397  & $3^*$ & $1^*$ & $4.8\times10^{30}$ & $0\degree - 25.8\degree$ & 0.83  \\
2.79 & 4.21 & 3 & 1 & $2.9\times10^{30}$ & $45.6\degree - 53.1\degree$ & 0.82 \\
2.23 & 0.101 & $3^*$ & $1^*$ & $4.5\times10^{30}$ & $45.6\degree - 53.1\degree$ & 0.78 \\
 - & - & 3 & 1 & $4.1\times10^{30}$ & $45.6\degree - 53.1\degree$ & 2.08 \\
2.49  & 0.808  & 3 & 1 & $3.2\times10^{30}$ & $84.3\degree - 90\degree$  & 1.37 \\
3.53  & 403.0  & $3^*$ & $1^*$ & $1.6\times10^{31}$ & $84.3\degree - 90\degree$  & 0.79  \\ 
3.03 & 19.3  & 3 & 2 & $1.8\times10^{30}$  & $0\degree - 25.8\degree$ & 0.77 \\
2.54 & 1.03 & 3 & 2 & $1.1\times10^{30}$ & $45.6\degree - 53.1\degree$ & 1.09 \\
3.13 & 35.5 & $3^\dagger$ & $2^\dagger$ & $1.8\times10^{30}$ & $45.6\degree - 53.1\degree$ & 0.89 \\
3.21 & 47.6 & $3^*$ & $2^*$ & $3.1\times10^{30}$ & $45.6\degree - 53.1\degree$ & 0.75 \\
3.10 & 2.73 & 3 & 5 & $1.2\times10^{30}$ & $0\degree - 25.8\degree$ & 0.83 \\
2.63 & 1.91 & 3 & 5 & $8.3\times10^{29}$ & $45.6\degree - 53.1\degree$ & 1.22  \\ 
2.66 & 1.66 & $3^*$ & $5^*$ & $2.3\times10^{30}$ & $45.6\degree - 53.1\degree$ & 0.74  \\
 - & - & $3^*$ & $5^*$ & $3.1\times10^{30}$ & $45.6\degree - 53.1\degree$ & 1.50  \\
2.10 & 0.0811 & 3 & 5 & $4.6\times10^{29}$ & $84.3\degree - 90\degree$  & 1.72  \\
 2.75 & 3.58 & $3^*$ & $5^*$ & $7.6\times10^{30}$ & $84.3\degree - 90\degree$  & 0.92  \\
 - & - & $3^*$ & $5^*$ & $1.1\times10^{31}$ & $84.3\degree - 90\degree$  & 2.17  \\
3.17 & 48.0 & 3 & 8 & $1.2\times10^{30}$ & $0\degree - 25.8\degree$ & 0.85 \\
2.57 & 1.34 & 3 & 8 & $7.6\times10^{29}$ & $45.6\degree - 53.1\degree$ & 1.27 \\
2.03 & 0.0562 & 3 & 8 & $2.8\times10^{30}$ & $84.3\degree - 90\degree$ & 1.74 \\
2.90  & 7.54 & 4 & 5 &  $3.2\times10^{31}$ & $0\degree - 25.8\degree$ & 0.75 \\
2.79  & 4.64  & 4 & 5 &  $2.3\times10^{31}$ & $45.6\degree - 53.1\degree$ & 0.96 \\
2.32  & 0.288  & 4 & 5 & $2.3\times10^{31}$ & $84.3\degree - 90\degree$ & 1.55 \\
3.54  & 396  & $4^*$ & $8^*$ &  $6.1\times10^{31}$ & $45.6\degree - 53.1\degree$ & 0.75 \\
2.89  & 9.16  & 4 & 8 &  $2.5 \times 10^{31}$ & $45.6\degree - 53.1\degree$ & 1.05 \\
2.93  & 10.5  & $4^\dagger$ & $8^\dagger$ &  $3.3 \times 10^{31}$ & $45.6\degree - 53.1\degree$ & 0.77 \\
2.86  & 6.6  & $4^\dagger$ & $8^\dagger$ & $7.6\times10^{31}$ & $84.3\degree - 90\degree$ & 1.19 \\
2.33 & 0.223 & 5 & 5 & $5.1\times10^{32}$ & $0\degree - 25.8\degree$ & 0.81 \\
- & - & 5 & 5 & $8.7\times 10^{32}$ & $0\degree - 25.8\degree$ & 1.80 \\
2.36 & 0.347 & 5 & 5 & $5.6\times10^{32}$ & $84.3\degree - 90\degree$ & 1.33 \\
- & - & 5 & 5 & $1.1\times10^{33}$ & $84.3\degree - 90\degree$ & 2.79 \\
\hline
\end{tabular}
\end{table}

\section{Discussion and Conclusions} 
      \label{sconc} 

FLUKA can be a useful tool for modelling ion transport, pion production and the observable consequences in flares. The directionality of the secondary products is modelled, as well as the transport of the escaping photons \citep[cf.][]{2018ApJ...869..182S}. The aim of the FLUKA authors to provide a consistent treatment of all relevant processes means that it also includes, e.g., the contribution of ``knock-on" electrons, primary proton bremsstrahlung, photons, and other secondaries produced via the decay of heavier, strange mesons and baryons (e.g. $K$ mesons, $\Lambda$'s; these are produced for ion energies $\gtrsim 2.5$~GeV but make only minor contributions to the observed fluxes), etc.

Our discussion of the 12 June 2010 flare shows that the spectrum can constrain the primary ion energy and angular distribution. In the case of this particular flare, too many primary ions directed towards the line of sight results in a spectrum that is harder than that observed. The best fits avoid this either by directing primary ions mostly downwards (or at least away from the observer if field lines are not vertical) or by viewing the flare from close to the \emph{effective} downward vertical direction and minimising the maximum primary ion energy. The capacity to model the angular dependence of the emission, as provided by FLUKA or similar codes, is clearly essential to making such statements. In particular angle-averaged spectra resemble those seen when the emitting region is viewed from a large angle to the vertical: generally harder, extending to higher photon energies, with a less pronounced $\pi^0$ feature (cf. Fig.~\ref{d3e58}). 

In reality we expect a more complex situation in which the direction of the field as well as its strength and gradient vary both with height and laterally in the chromosphere \citep{2020MNRAS.491.4852H}. The magnetic mirror force will change the velocity vectors of ions as they slow down in the source, in a way not included in our FLUKA simulations. Apart from viewing-angle effects, ions in long-duration events may be trapped in low density regions. In such a situation $\pi^0$ decay radiation will still be produced instantaneously but there will be less accompanying radiation from secondary $e^{\pm}$ because their lifetimes are much longer in low density conditions, and/or they are suppressed by synchrotron energy losses \citep{1987ApJS...63..721M}. Clearly our FLUKA simulations yield constraints that are useful in their own right but how to satisfy these constraints may need further consideration employing a detailed model of the chromospheric and photospheric magnetic field.

We concentrated on a short event less likely to involve radiation from particles trapped in low density regions. The forthcoming Fermi-LAT Solar Flare Catalog (M. Pesce-Rollins, private communication) emphasises ``impulsive" or ``gradual" as a fundamental distinction between high-energy events. It will be interesting to see if the data analysis approach here, applied across many of these events, highlights systematic differences between these two basic classes.

Subsequent work will obtain best fits to observed spectra for many events, aiming to exploit the diagnostic potential discussed here, as well as considering other sorts of emission from secondary particles, e.g. (gyro)synchrotron radiation that should peak in the THz range. This has important consequences for diagnostics using high frequency observatories such as the Atacama Large Millimeter Array (ALMA) \citep{2017IAUS..328..120T,2016SSRv..200....1W}.

\begin{acks}[Acknowledgements]
This research was partially supported by the Brazilian agency FAPESP (contracts 2009/18386-7, 2013/24155-3, 2017/13282-5 and support for a sabbatical visit by ALM), by an STFC Consolidated Grant and by the Royal Society Newton Fund. The work also benefitted from the EU FP7 IRSES Grant 295272, 'RadioSun'. GGC acknowledges support from the Brazilian Science Agency CNPq through the Productivity in Science Fellowship.  JT acknowledges FAPESP support (Grant 	
2016/23428-4). ALM thanks his co-authors and their CRAAM colleagues for warm hospitality on several visits to Mackenzie University and acknowledges particularly the support over several years of the late P. Kaufmann. H Hudson and P Sim\~{o}es commented helpfully on an earlier draft.  We thank the authors of FLUKA and acknowledge the support provided via the fluka-discuss mailing list. This work benefited from discussions at the ISSI International Team, "Energetic Ions: the Elusive Component of Solar Flares", and from further advice on LAT data from M Pesce-Rollins and N Omodei.
\end{acks}

\section*{Disclosure of Potential Conflicts of Interest}

The authors declare that they have no conflicts of interest.

\appendix

Here we note some features of FLUKA that should be borne in mind when using it for the solar atmosphere, reiterating some points made in \cite{2019SoPh..294..103T} and adding a few further comments.

\begin{itemize}
\item FLUKA treats densities $< 10^{-10}$ g cm$^{-3}$ as vacuum. This is the density of the chromosphere at a height of $\approx 600$ km in the VAL-C \citep{1981ApJS...45..635V} model, so a literal model of the solar atmosphere is not feasible. This is not a major problem for present purposes since the radiative ouput from all of a thick target, or from a layer of fixed column density, is what is important, irrespective of its physical size - as long as this is much larger than the muon decay length, and as long as we also pay attention to the following. 
\item FLUKA models the slowing down of ions assuming a neutral medium. The Coulomb logarithm may be significantly larger (typically by a factor of up to three) in an ionised medium \citep{1956PThPh..16..139H,2003A&A...409..745M,2015SoPh..290.2809T}. Most previous  calculations of solar pion decay radiation employ neutral medium stopping rates, appropriate to the deep atmosphere \citep[e.g.][]{1987ApJS...63..721M,2010ApJ...721.1174T}. We continue with this practice here.
\item However, ionisation potentials used to calculate the Bethe-Bloch stopping rate (\citealp{2014ChPhC..38i0001O}, Section 32) are by default those appropriate to molecules, e.g. H$_2$, O$_2$, etc., slightly different from the atomic values. Different values appropriate to single atoms may be inserted by hand. Since the ionisation potential appears in the argument of the logarithm the difference is slight in any case. Attempting to model energy loss in an ionised medium by substituting $h\nu_p$ for an atomic ionisation potential is too substantial a departure for FLUKA to accommodate, however.
\item FLUKA includes the modifications to the slowing-down rate that result from the collective response of a neutral medium \citep[e.g.][]{2002NIMPB.187..285W,2014ChPhC..38i0001O}. The consequence is that the yield of secondaries from a specified target depends not only on the total column depth, but slightly on the actual density encountered. This correction becomes most important at the highest energies but is only ever of order unity. We checked that our results did not depend on density to a degree that would be important for interpreting observed fluxes. However, the parametric (Sternheimer) form used to represent this effect breaks down, giving runtime errors, at densities slightly higher than the lower limit of $10^{-10}$ g cm$^{-3}$ mentioned above; another factor that discourages a literal model of the solar atmosphere.   
\item Sometimes we want information on secondaries resulting from unlikely events (e.g. Figure~\ref{beamvsiso}). Then statistically reliable spectra are obtained more easily by activating ``biasing'' in FLUKA, specifically an artificial reduction in the mean free path for hadronic interactions. 

\end{itemize}


\bibliographystyle{spr-mp-sola}

\bibliography{fluka}  

\IfFileExists{\jobname.bbl}{} {\typeout{}
\typeout{****************************************************}
\typeout{****************************************************}
\typeout{** Please run "bibtex \jobname" to obtain} \typeout{**
the bibliography and then re-run LaTeX} \typeout{** twice to fix
the references !}
\typeout{****************************************************}
\typeout{****************************************************}
\typeout{}}

\end{article} 

\end{document}